\documentclass[twocolumn,pra,aps,showpacs]{revtex4}
\usepackage{epsfig,hyperref}
\def\be{\begin{equation}}
\def\ee{\end{equation}}
\def\ba{\begin{eqnarray}}
\def\ea{\end{eqnarray}}
\def\haa{\hat{a}}
\def\hac{\hat{a}^{\dagger}}
\def\hba{\hat{b}}
\def\hbc{\hat{b}^{\dagger}}

\makeatletter
\@addtoreset{equation}{section}
\makeatother

\begin{document}
\title{Superfluidity of Bose-Einstein Condensate in An Optical Lattice:  
Landau-Zener Tunneling and Dynamical Instability}
\author{Biao Wu}
\affiliation{Department of Physics, The
University of Texas, Austin, Texas 78712}
\affiliation{Condensed Matter Science Division, Oak Ridge National Laboratory,
Tennessee 37831-6032, USA}
\author{Qian Niu}
\affiliation{Department of Physics, The
University of Texas, Austin, Texas 78712, USA}

\date{July 2, 2003}

\begin{abstract}
Superflow of Bose-Einstein condensate in an  optical lattice is represented by
a Bloch wave, a plane  wave with  periodic  modulation of the amplitude.    We
review the theoretical results on the interaction effects  in the energy
dispersion of the Bloch waves and in the linear stability  of such waves.   For
sufficiently strong repulsion between the atoms, the lowest Bloch band develops
a loop at  the edge of the Brillouin zone, with the dramatic consequence of  a
finite probability of Landau-Zener tunneling even in the limit of a vanishing
external force.    Superfluidity can exist in the central region of the
Brillouin zone in the presence of  a repulsive interaction, beyond which Landau
instability takes place where the system can lower its energy by making
transition into states with smaller Bloch wavenumbers.    In the outer part of
the region of Landau instability, the Bloch waves are also dynamically unstable
in the sense that a small initial deviation grows exponentially  in time.   In
the inner region of Landau instability,  a Bloch wave  is  dynamically stable
in the absence of persistent external perturbations.     
Experimental implications of our findings will be discussed.  
\end{abstract}
\maketitle
\section{Introduction}
As one of the most remarkable macroscopic quantum phenomena, superfluidity
has attracted enormous attention since its discovery\cite{superflu,Lifshitz,fm}. 
The realization of Bose-Einstein condensation in dilute atomic gases\cite{bec}
has provided physicists a new fertile ground for exploring many aspects of
this fascinating phenomenon, including frictionless current\cite{Raman} and
vortices \cite{vortex}. In particular, there has been increasing interests 
and efforts to study superfluidity and related phenomena in the system 
of a Bose-Einstein condensate (BEC) in an optical lattice, such
as Landau-Zener tunneling\cite{choi,loop,zobay,morsch}, Josephson 
effect\cite{josephson}, and dynamical instability\cite{burger,wn,bronski}. 
Recently, quantum phase transition between superfluidity  and 
Mott-insulator\cite{mott_ex,mott_th} was observed in such a system.

Superflow of BEC in an  optical lattice is represented 
by a Bloch wave, which can be regarded as a plane  wave modulated by the 
periodic potential. Bloch waves and Bloch bands are the basic language and 
concepts in periodic systems. They are also essential to the study
of superfluidity and many of its related phenomena in this periodic BEC system.

One interesting phenomenon is the tunneling between Bloch bands 
at the edge of the Brillouin zone. Previously, people studied 
experimentally this inter-band tunneling by dragging cold atoms 
with accelerated optical lattices\cite{Bha}. The cold atoms are very dilute 
and the interaction between them can be ignored entirely. It is
then curious to know how the tunneling will change if the cold atoms
are replaced with a BEC, where the atomic density is relatively large
and the interaction between atoms can no longer be neglected. 
Our study shows that the interaction will increase the tunneling 
probability in general\cite{loop,tnlz}. When the repulsive interaction is 
over a critical value, a loop appears in the lowest Bloch band,
resulting in non-zero tunneling probability even in the adiabatic 
limit\cite{loop,tnlz,wdn,smith,smith2}.
This breakdown of adiabaticity is the result of superfluidity and can be viewed
as a hysteresis phenomenon\cite{loop,mueller}.

The most dramatic manifestation of superfluidity is that the boson system can 
maintain its current speed in a very tight space, such as a narrow capillary.
Landau proposed a criterion for superfluidity\cite{Lifshitz,fm}: 
if the current moves slower than sound, it experiences no viscosity; otherwise,
the system suffers an instability and loses its speed. In a homogeneous
BEC, the sound speed is proportional to the square root 
of the BEC density. It is not clear 
how superfluidity can be defined and studied for a BEC in an optical lattice. 
Our method is to examine the energy dispersion of Bloch waves.
When the Bloch waves are energy minima of the system, they
represent superflows; when the Bloch waves are energy saddle points,
they suffer Landau instability: external disturbances can render
the system to emit phonons thus reduce its speed.

The BEC Bloch waves can be dynamically unstable, that is, the system diverges 
away from the original Bloch state upon small instantaneous 
perturbation\cite{wn}. This instability does not exist either for
a BEC in a free space or for a free particle in a periodic lattice; it 
happens only when there are both interaction and periodic lattice. The
dynamical instability has been observed experimentally\cite{burger,wnprl}, 
and further confirmed by other theoretical studies and experimental 
observations\cite{smith2,smerzi,bishop}. This dynamical instability is 
quite a general phenomenon and widely exists in other systems, such as a 
BEC confined in a ring\cite{garay,sinha}. Its general implication and 
relation to superfluidity will be discussed in this review.
 
We will discuss these phenomena in detail and review many interesting results for
the quasi-one dimensional case, where the optical lattice is created by two 
counter-propagating off-resonance lasers while the lateral motion of the BEC 
can be either ignored or confined\cite{choi,morsch,1d}. In Section II, a 
brief introduction of the mean-field theory of BEC systems is given for 
the sake of completeness and 
introduction of the notations. In Section III, we define Bloch waves and Bloch 
bands for BECs in optical lattices and present some general results and the 
numerical methods to find them. In Section IV, we simplify the system to a 
two-level model to study the tunneling between Bloch bands. We point out
the tunneling in the adiabatic limit is related to a general problem,
adiabatic evolution of nonlinear quantum states. In Section V, the 
Landau instability and dynamical instability of the Bloch waves are studied. 
Other approaches to superfluidity and experimental observation and general 
implication of dynamical instability are discussed. Finally in Section VI, 
we summarize the paper.

\section{Mean-field Theory of BEC systems}
Even though the BEC is quite dense compared to the cold atoms, it is still
very dilute with typical densities at $10^{-20}$m$^{-3}$, which is thousands of
times more dilute than the air. Due to this diluteness, along with the extreme
low temperature, the interaction
between atoms can be characterized by the $s$-wave scattering and modeled 
with a $\delta$-function. When the number of atoms is large, the BEC system
is very well described by the mean-field theory given by the 
celebrated Gross-Pitaevskii equation\cite{Gross}
\be\label{eq:nls0}
i\hbar{\partial \psi\over\partial t}=-{\hbar^2\over 2m}{\partial^2\psi\over
\partial x^2}+V(x)\,\psi+{4\pi\hbar^2a_s\over m}|\psi|^2\psi\,,
\ee
where $m$ is the atomic mass, $a_s$ is the $s$-wave scattering length, and
$V(x)$ is  the external potential imposed on the system. In our case, the
potential is the optical lattice created by two laser beams and is given by
\be
V(x)=V_0\cos(2k_L x)\,,
\ee
where $V_0$ is proportional to the laser intensity and $k_L$ is the wave 
number of the laser. 

For simplicity, we will instead use the following dimensionless 
Gross-Pitaevskii (GP) equation 
\be\label{eq:nls}
i{\partial\psi \over\partial t}=-{1\over 2}{\partial^2\psi\over \partial x^2}
+v\cos(x)\,\psi+c|\psi|^2\psi\,.
\ee
In the above equation, all the variables are scaled to be dimensionless
with the system's basic parameters: the strength of the periodic potential
$v$ is in units of ${4\hbar^2 k_L^2\over m}$, the wave function $\psi$ in
units of $\sqrt{n_0}$ where $n_0$ is the averaged BEC density, $x$ in units 
of ${1\over 2k_L}$, $t$ in units of ${m\over 4\hbar k_L^2}$.  
The coupling constant $c={\pi n_0 a_s\over k_L^2}$.

This system can also be regarded as a Hamiltonian
system governed by the grand canonical Hamiltonian
\be\label{eq:h}
H=\!\int\!{\rm d}x~\Big\{\psi^*\Big(-{1\over 2}
{\partial^2\over \partial x^2}+v\cos(x)\Big)\psi+{c\over 2}|\psi|^4-\mu|\psi|^2\Big\},
\ee
where $\mu$ is the chemical potential. The GP equation (\ref{eq:nls}) can
be obtained by variation of the Hamiltonian, $i\partial \psi/\partial t=
\delta H/\delta \psi^*$.

\section{Bloch Waves and Bloch Bands}
Among all possible solutions of Eq.(\ref{eq:nls}), there are states
which extremize the Hamiltonian (\ref{eq:h}) and hence satisfy the 
time-independent  Gross-Pitaevskii equation
\be\label{eq:nlsi}
-{1\over 2}{d^2\psi\over d x^2}
+c|\psi|^2\psi+v\cos(x)\,\psi=\mu\,\psi\,.
\ee
We call these extremum states nonlinear eigenstates, which are also called 
nonlinear coherent modes elsewhere\cite{Yukalov}; accordingly,
$\mu$ are nonlinear eigenvalues. Bloch waves
are the nonlinear eigenstates of the following form
\be\label{eq:bloch}
\psi(x)=e^{ikx}\phi_k(x)\,,
\ee
where $\phi_k(x)$ is a periodic function of period $2\pi$ and
$k$ is the Bloch wave number. In particular, from Eq.(\ref{eq:nlsi})
we have the following equation for each Bloch wave state
$\phi_k$ 
\be\label{eq:nlsp}
-{1\over 2}({d\over d x}+ik)^2\phi_k
+c|\phi_k|^2\phi_k+v\cos(x)\,\phi_k=\mu(k)\,\phi_k\,.
\ee
The set of eigenvalues $\mu(k)$ then form Bloch bands.

When $v=0$, the plane waves are the solutions of Eq.(\ref{eq:nlsi}), and 
they represent a BEC flow with speed of $k$. As is well known, when the speed 
is smaller than the speed of sound, $k<\sqrt{c}$, it is a superflow;
when $k>\sqrt{c}$, it develops a Landau instability and loses the 
superfluidity. Bloch waves are the counterpart of these plane waves in a 
periodic system. Whether these Bloch waves represent superflows or not 
is determined by the energy dispersion of these Bloch waves as we will show later.

\begin{figure}[!htb]
\begin{center}
\resizebox*{8cm}{6cm}{\includegraphics*{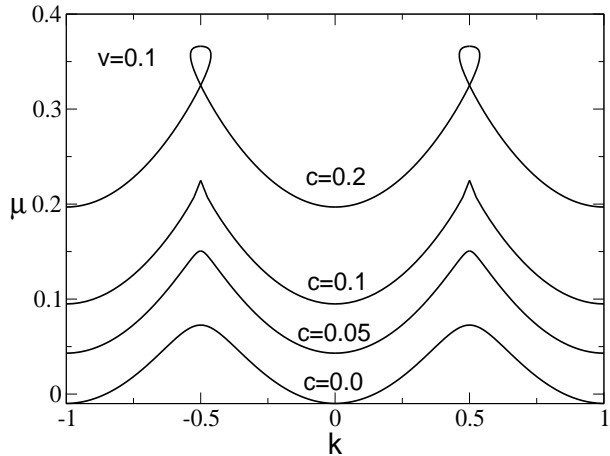}}
\end{center}
\caption{Lowest Bloch bands at $v=0.1$ for $c=0.0$, $c=0.05$, $c=0.1$, 
and $c=0.2$ (from bottom to top). As $c$ increases, the tip of the Bloch band 
turns from round to sharp at the critical value $c=v$, followed by the 
emergence of a loop.
}
\label{fig:bband}
\end{figure}

In the linear case, $c=0$, all eigenstates are
Bloch waves\cite{Mermin}. The situation is very different for the periodic 
nonlinear system, where it is possible to have non-Bloch wave eigenstates. 
Furthermore, there are nonlinear Bloch waves that do not have their linear 
counterparts. Here is a beautiful example\cite{bronski,wdn}. Eq.(\ref{eq:nlsi})
has an exact solution of Bloch wave at the Brillouin zone edge $k=1/2$,
\be
\label{eq:exact}
\psi={\sqrt{c+v}+\sqrt{c-v}\over 2\sqrt{c}}e^{i{x\over 2}}-
{\sqrt{c+v}-\sqrt{c-v}\over 2\sqrt{c}}e^{-i{x\over 2}}\,.
\ee
This Bloch wave state only exists when $c\ge v$ so that it has no linear counterpart. 
This shows that there are some ``extra'' 
nonlinear Bloch waves when the nonlinearity is strong enough. 
Their corresponding ``extra'' eigenvalues should make 
the nonlinear Bloch band look very different from the linear Bloch band. 
This is indeed the case: our numerical results show a loop structure
in the lowest Bloch band when $c>v$ as seen in Fig.\ref{fig:bband}.
This loop structure was first found in Ref.\cite{loop} with a two-mode
approximation, then confirmed in Ref.\cite{smith,smith2}. 
In particular, in Ref.\cite{smith2}, a loop is also found for
the second band at the Brillouin zone center.

This loop structure is a consequence of superfluidity. One clear indication
is that the exact Bloch wave at the Brillouin zone edge, Eq.(\ref{eq:exact}), 
represents a flow of non-zero speed, $\sqrt{c^2-v^2}/2c$\cite{wdn}. This is quite
remarkable if we recall that all the Bloch waves at the zone edge carry
no currents in the linear case. For a free particle, the flow $e^{ix/2}$
is completely stopped due to first-order Bragg scattering by the periodic
potential, leading a zero-current Bloch waves at the edge, $k=1/2$. However,
when the superfluidity gets stronger, that is, when $c>v$, the flow $e^{ix/2}$
can no longer be stopped by Bragg scattering, yielding a zone-edge Bloch wave
that carries a current.  The connection between superfluidity and the loop
structure is discussed in-depth in the context of hysteresis by 
Mueller\cite{mueller}.

These Bloch states can be prepared experimentally, at least for
the ones in the lowest Bloch band by adiabatic control. For a trapped 
cigar-shaped BEC, we slowly turn on an optical lattice along its 
longitudinal direction; we then have a BEC in the Bloch state at the center
of the Brillouin zone. Other Bloch states can be achieved by accelerating
the lattice for a certain amount of time. These experimental techniques
of adiabatic turning-on and accelerating optical lattices have been demonstrated 
successfully with either cold atoms\cite{Bha,bo} or BECs\cite{morsch}. 

There are several numerical methods to find the Bloch wave $\phi_k$. 
The method described in Ref.\cite{wn} is to expand $\phi_k$ in Fourier series
and minimize the Hamiltonian (\ref{eq:h}) in the space expanded by Fourier
coefficients. Another one is used in Ref.\cite{smith}, where Eq.(\ref{eq:nlsi}) 
is first solved for different values of $\mu$ and $k$ then the Bloch 
band $\mu(k)$ is found by interpolation. 

A much better method is the following. 
One still starts with the expansion of $\phi_k$ in Fourier series,
\be
\phi_k(x)=\sum_{n=-N}^{N}a_i e^{inx}\,,
\ee
where $N$ is the cut-off. With the substitution of the above equation 
into Eq.(\ref{eq:nlsp}), one obtains $2N+1$ equalities for the coefficients
of each Fourier term $e^{inx}$,
\be
f_n(a_0,a_{\pm 1},\cdots,a_{\pm N},\mu)=0\,.
\ee
Finally, the Bloch wave is found by minimizing the following sum
\be
S=\sum_{n=-N}^N f_n^2\,.
\ee
The results in Fig.\ref{fig:bband} are obtained with this method by using $N=10$.

\section{Tunneling Between Bloch Bands}
Consider the scenario that we drag the BEC across the entire
Brillouin zone with an accelerating lattice. If the BEC is initially
in a Bloch state belonging to the lowest Bloch band and the acceleration 
is small enough, the system will stay in the band and undergo Bloch 
oscillations\cite{choi,morsch}. As one increases the acceleration, the 
BEC will have increasing chance tunneling into the upper band at the edge 
of the Brillouin zone, where the band gap is the smallest. Without interaction,
this tunneling is nothing but the famous Landau-Zener tunneling\cite{lz}, 
a well understood phenomenon.  With the interaction in our case,  
the tunneling behavior is strongly modified, in particular, by
the loop structure in the Bloch band as we will see later.

In an accelerating lattice, the system is described by 
\be\label{eq:nlsa}
i{\partial \over\partial t}\psi=-{1\over 2}
({\partial\over\partial x}+i\alpha t)^2\psi+v\cos(x)\psi+c|\psi|^2\psi\,,
\ee
where $\alpha$ is the acceleration. Even in the linear case, $c=0$,
the above equation is difficult to analyze. To reduce
the mathematical complexity,  we will approximate it with a 
two-level model without losing the essential physics. 

\subsection{Two-level Model}
The tunneling mainly occurs around the edge of the Brillouin zone, $k=1/2$, 
where the wave function is dominated by two plane wave 
components. We zoom in on this region and write 
\begin{equation}
\phi(x, t)=a(t)e^{ikx}+b(t)e^{i(k-1)x},
\end{equation}
where $|a|^2+|b|^2=1$. Following the simple algebra described 
in Ref.\cite{loop} by substituting the above two-mode wave function  into 
Eq.(\ref{eq:nlsa}), we obtain a two-level nonlinear Schr\"odinger equation
\begin{equation}
\label{eq:twolevel}
i{\partial\over \partial t}\pmatrix{ a\cr  b}= 
H(\gamma)\pmatrix{a\cr b}\,,
\end{equation}
where 
\begin{equation}
H(\gamma)={1\over 2}\pmatrix{\gamma + c\,\kappa & v\cr 
v &-\gamma -c\,\kappa}\,,
\end{equation}
where $\gamma=\alpha t$ and $\kappa=|b|^2-|a|^2$. 

\begin{figure}[!htb]
\begin{center}
\resizebox*{8cm}{6cm}{\includegraphics*{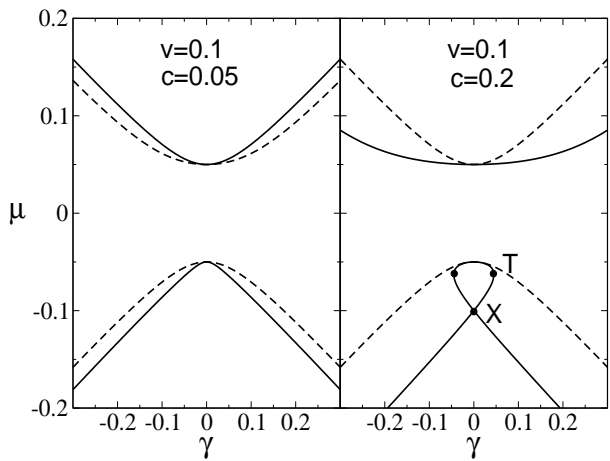}}
\end{center}
\caption{Adiabatic energy levels for $c=0.05$ (left) and for $c=0.2$ (right)
at $v=0.1$. For comparison, the levels for the linear case, $c=0.0$,
are plotted in dashed lines.}
\label{fig:loop2}
\end{figure}

However simplified it may appear, this two-level model captures the
essence of our BEC system as it reproduces the looped Bloch band
(see  Fig.\ref{fig:loop2}). In this simple model, the ``Bloch bands''
are obtained by finding the eigenvalues of $H(\gamma)$,
\be
H(\gamma)\pmatrix{a\cr b}=\mu(\gamma)\pmatrix{a\cr b}\,.
\ee
The eigenvalues $\mu(\gamma)$ are also called adiabatic energy levels.
How the levels change with the coupling strength $c$ is shown in Fig.\ref{fig:loop2},
where we again see the appearance of the loop for $c>v$ in the lower energy level.

For this two-level model, the original tunneling problem 
in the BEC system can be restated as follows:
if the system starts in the lower level, as the energy bias
$\gamma$ changes with the sweeping rate $\alpha$ from the far negative end
to the far positive end, what is probability that the system ends up in the upper
level?

When $c=0$, this is precisely the well-known Landau-Zener model, 
which can be solved exactly. The tunneling probability is\cite{lz}
\begin{equation}
\label{eq:r_lz}
r_0=\exp(-{\pi v^2\over 2\alpha})\,. 
\end{equation} 
It is clear from this formula that, in the adiabatic limit $\alpha\rightarrow 0$,  
the tunneling probability is zero, that is, the system stays in the 
lower level throughout the entire process. This is nothing but a special case 
of the quantum adiabatic theorem\cite{landau}: in the adiabatic change of a parameter 
the system starting in an eigenstate stays in the same instantaneous eigenstate.

When $c>v$, this adiabaticity is broken down by the loop structure:
Suppose we start with a state on the lower adiabatic level, 
and move it up along the branch by changing $\gamma$ 
so slowly such that little tunneling to the upper level is generated. 
After passing the crossing point $X$, the state remains
in the course moving up in energy until hitting the terminal point $T$, 
where it has no way to go any further except to jump to 
the upper and lower levels. This leads to a nonzero probability  
tunneling into the upper level in the adiabatic limit.  

\begin{figure}[!htb]
\begin{center}
\resizebox*{8cm}{7.5cm}{\includegraphics*{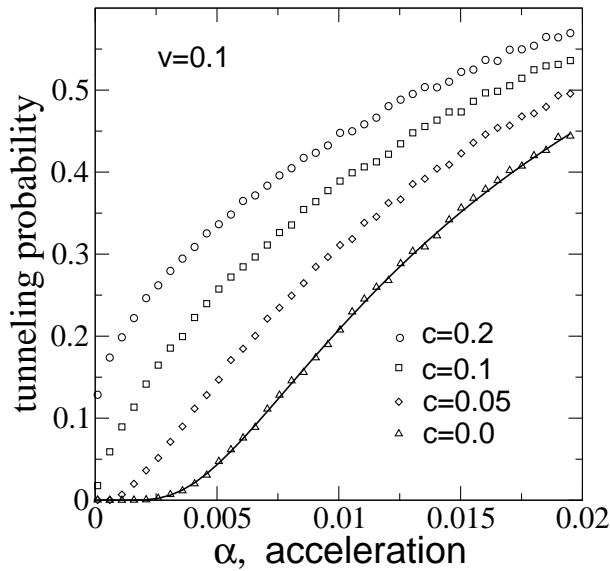}}
\end{center}
\caption{Numerical results of tunneling probability as a function 
of $\alpha$ at $v=0.1$ for different values of $c$. For comparison, 
the exact result Eq.(\ref{eq:r_lz}) for the linear case $c=0$ is
plotted in solid line. The small wiggles are not physical and they
are resulted from the numerical errors of our computer codes.
}
\label{fig:rate}
\end{figure}

This qualitative analysis is corroborated by our numerical integration
of Eq.(\ref{eq:twolevel}). The tunneling probabilities as a function
of the sweeping rate $\alpha$ are plotted in Fig.\ref{fig:rate} for different
values of $c$ at $v=0.1$. In general, the tunneling probability
is enhanced due to the interaction. However, its dependence on
$\alpha$ changes dramatically with the increasing interaction strength $c$.
When $c<v$, the tunneling probability
approaches zero in the adiabatic limit, $\alpha\rightarrow 0$. It appears
that the approaching is exponential, as supported by a recent analysis\cite{tnlz}.
At the critical point, $c=v$, the tunneling probability
still turns to zero as the sweeping slows down, however, not exponentially.

When $c$ is over the critical value, it becomes apparent in Fig.\ref{fig:rate}
that the  tunneling curve tends to intersect the vertical axis at a finite
value, indicating the tunneling probability is not zero at the 
adiabatic limit, $\alpha\rightarrow 0$. Of course, one can never be absolutely 
sure that the tunneling probability at $\alpha\rightarrow 0$ is not zero 
by numerical calculation since there is always a limit on the smallness
of $\alpha$ in a computer code. This uncertainty is put to an end in
Ref.\cite{tnlz}, where this nonzero adiabatic tunneling probability
is obtained analytically.

One can also directly integrate numerically the original Schr\"odinger
equation Eq.(\ref{eq:nlsa}) to find out the tunneling probability and
how it depends on the acceleration $\alpha$. This is done in Ref.\cite{wuchoi},
where one sees a similar crossover from the exponential to non-exponential
dependence of the tunneling probability on $\alpha$. We argue that 
this crossover behavior can serve as an indirect evidence of the existence
of the loop structure in the Bloch band. We believe that this crossover
should be observable in an experiment similar to the one conducted in 
Ref.\cite{morsch}. 

The nonlinear Landau-Zener tunneling studied here is likely as general as 
the linear counterpart and should exist in many physical systems. We found
one example in molecular magnets\cite{mm}, where the jamming effect among
interacting molecular spins is the result of nonlinear Landau-Zener tunneling.

\subsection{Nonlinear quantum adiabatic theorem}
The nonlinear two-level model (\ref{eq:twolevel}) is a very good
example of a general problem: how a nonlinear quantum system, such
as a BEC system, responses to the change of a system parameter.
Of particular interest is that the parameter change is adiabatic since
the adiabatic control is an important tool in preparing a quantum system 
(linear or nonlinear) into a desired quantum state \cite{qac}.   
For a linear quantum system, the answer is the quantum adiabatic theorem,
which can be found in a standard textbook\cite{landau}. The theorem states that 
if the system is initially in a non-degenerate eigenstate, it stays in the same
eigenstate as the parameter changes adiabatically. 

On the other hand, the answer for nonlinear quantum systems is not
as straightforward. The nonlinearity proves to be a big obstacle for
generalizing the linear quantum adiabatic theorem\cite{band}.
Very non-trivial modifications to the linear case are expected as 
it is found in the nonlinear two-level model (\ref{eq:twolevel}) that 
tunneling between energy levels happens even 
in the adiabatic limit.

In a recent paper\cite{nqat}, we have successfully generalized the quantum
adiabatic theorem to nonlinear quantum systems by combining 
ideas from classical adiabatic dynamics and quantum geometric phases. 
It is found that the eigenstates correspond to fixed points of 
an equivalent classical problem. The adiabatic evolution of these
eigenstates depends on the nature of the corresponding fixed points. 
If the fixed point is elliptic, the system can stay in
the same eigenstate as in the linear case. Otherwise, 
when the fixed point is hyperbolic, the system will stray away 
from the eigenstate, leading to tunneling between different eigenstates. 
This adiabatic condition can also be specified in terms of Bogoliubov
spectrum. If the Bogoliubov spectrum is real, the adiabaticity
is kept; otherwise, it is broken.

\section{Superfluidity and Dynamical instability}
\subsection{Superfluidity and Landau instability}
A quantum Bose liquid or gas at very low temperatures possesses a very 
remarkable property known as superfluidity: its flows suffers no viscosity
through capillaries or other types of tight spaces when its speed is slower
than a critical value. This was discovered even before the conception
of Bose-Einstein condensation\cite{superflu}. Landau\cite{Lifshitz,fm} 
gave a simple explanation for this intriguing phenomenon. He argued that, 
a quantum current suffers friction only when the system's elementary excitation, 
phonon, can lower its energy. With the Galilean transformation,
he found that a quantum liquid flowing with a speed smaller than
the sound speed enjoys such frictionless transport. When it flows
with a speed larger than the sound speed, the quantum liquid can
lower its energy by emitting phonons, leading to the slowdown of its
speed. 

Such lose of superfluidity can also be regarded as an instability: the system
can no longer stay in its original state thus keep its flow 
against perturbation of the roughness of capillaries or other external disturbance.
We will call this instability Landau instability. Note that this Landau instability
is different from the Landau instability discussed in plasma physics
\cite{vlasov}.

\begin{figure}[!htb]
\begin{center}
\resizebox*{7cm}{6.6cm}{\includegraphics*{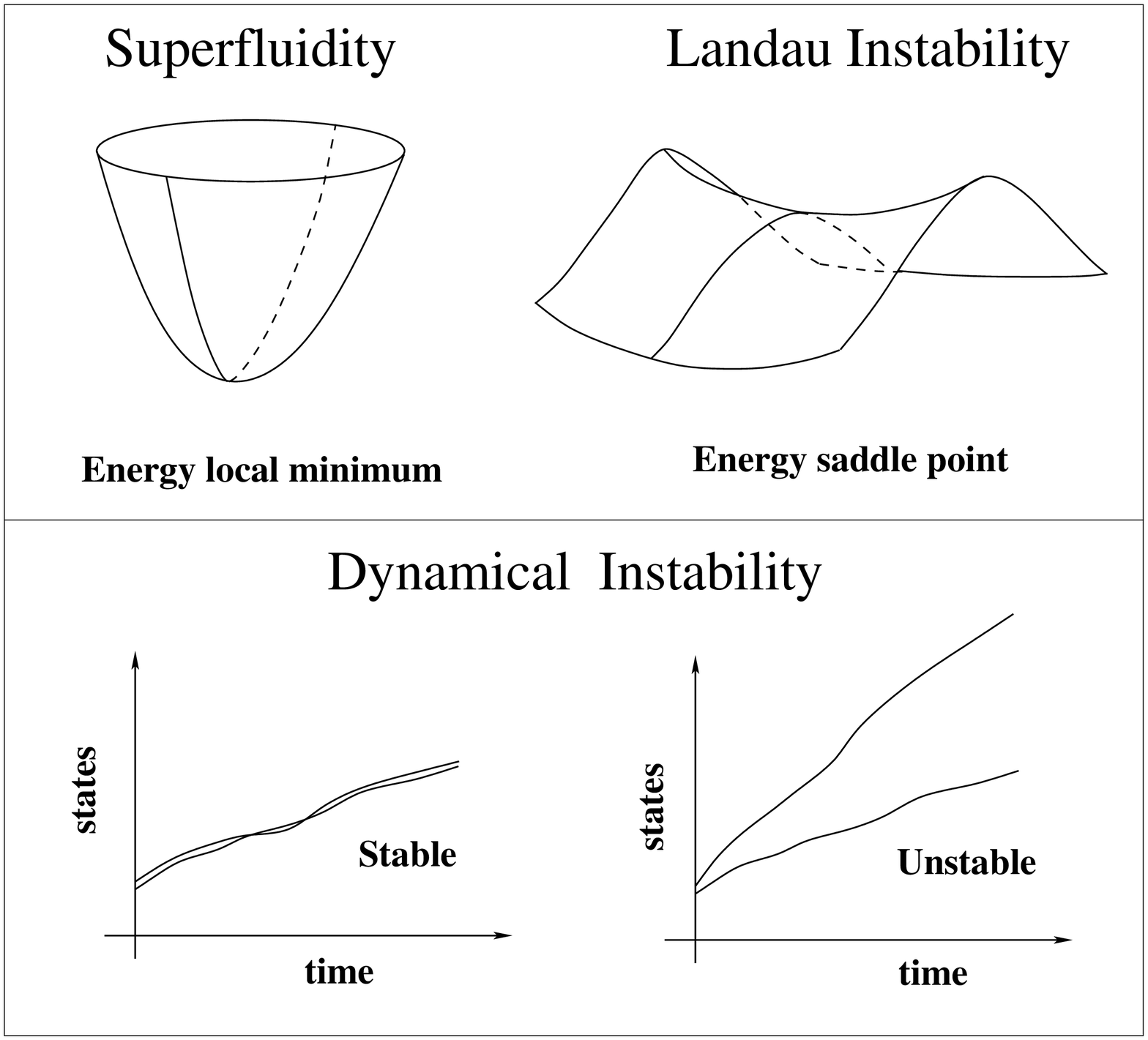}}
\end{center}
\caption{Schematic illustration of superfluidity, Landau instability,
and dynamical instability. A dynamically unstable state must be
a saddle point; a state that is an energy saddle point can be dynamically
stable.}
\label{fig:landyn}
\end{figure}

In our case, the study of superfluidity or Landau instability is
complicated by the presence of the optical lattice, which 
breaks the translational symmetry along its flow direction and 
modulates the BEC density. This complication is only mathematical.
The physics is still the same, that is, the Landau instability is 
determined by whether the elementary excitation around a BEC flow 
lowers the system's energy or not. If it does not, the BEC flow
is a local energy minimum of the system and it is a superflow.
Otherwise, the flow is an energy saddle point and it suffers Landau
instability (see Fig.\ref{fig:landyn}).

Whether these Bloch waves are local energy minima is determined by their 
energy dispersions. To calculate the dispersion, we perturb the system around a Bloch state
\be\label{eq:pert}
\psi(x)=e^{ikx}\Big[\phi_k(x)+\delta\phi_k(x)\Big]\,.
\ee
Due to the periodicity of our system and the Bloch waves, 
these perturbation can be decomposed  into different modes labeled by $q$,
\be 
\delta\phi_k(x,q) = u_k(x,q)e^{iqx}+v_{k}^*(x,q)e^{-iqx},
\ee
where $q$ is also a kind of Bloch wave number and ranges between $-1/2$ 
and $1/2$. The perturbation functions $u_k$ and $v_k$ are of periodicity 
of $2\pi$ in $x$.  By substituting the perturbed state (\ref{eq:pert}) into 
the Hamiltonian (\ref{eq:h})and neglecting
terms of orders higher than two,
we obtain a quadratic form of the energy deviation, which
is block diagonal in $q$. Each block is given by  
\be
\delta E_k=\int_{-\infty}^{\infty}{\rm d}x
\pmatrix{u_{k}^*,v_{k}^*}M_k(q)\pmatrix{u_{k} \cr v_{k}},
\label{eq:approx}
\ee
where 
\be
\label{eq:mmatrix}
M_k(q)=\pmatrix{{\mathcal L}(k+q) & c\phi_k^2 
\cr c\phi_k^{*2} & {\mathcal L}(-k+q)}, 
\ee
with
\be
{\mathcal L}(k)=-{1\over 2}({\partial \over \partial x} + ik)^2
+v\cos(x)-\mu+2c|\phi_k|^2 .
\ee
If $M_k(q)$ is positive definite for all $-1/2\le q\le 1/2$, the Bloch wave 
$\phi_k$ is a local minimum and represents a superflow . 
Otherwise, $\delta E_k$ can be negative for some $q$, indicating that
the Bloch wave is a saddle point and describes only a normal flow.

The special case $v=0$, when the optical lattice is turned off,
is the simple case considered by Landau\cite{Lifshitz}. In this case,
the Bloch state $\phi_k$ becomes a plane wave $e^{ikx}$ and 
the operator $M_k(q)$ becomes a $2\times 2$ matrix
\be
M_k(q)=\pmatrix{q^2/2+kq+c&c\cr c&q^2/2-kq+c}
\ee
whose eigenvalues are
\be\label{eq:lambda}
\lambda_{\pm}={q^2\over 2}+c\pm\sqrt{k^2q^2+c^2}.
\ee
Since $\lambda_{+}$ is always positive, the matrix $M_k(q)$ is  not
positive definite only when $\lambda_{-}\le 0$, or equivalently,
$|k|\ge \sqrt{q^2/4+c}$. It immediately follows that
the BEC flow $e^{ikx}$ becomes a saddle point when the flow speed exceeds
the sound speed, $|k|>\sqrt{c}$.  This recovers exactly the Landau condition 
for the breakdown of superfluidity \cite{Lifshitz}, which has recently 
been confirmed experimentally on BEC \cite{Raman}.   

\begin{figure}[!htb]
\begin{center}
\resizebox*{8.5cm}{11.5cm}{\includegraphics*{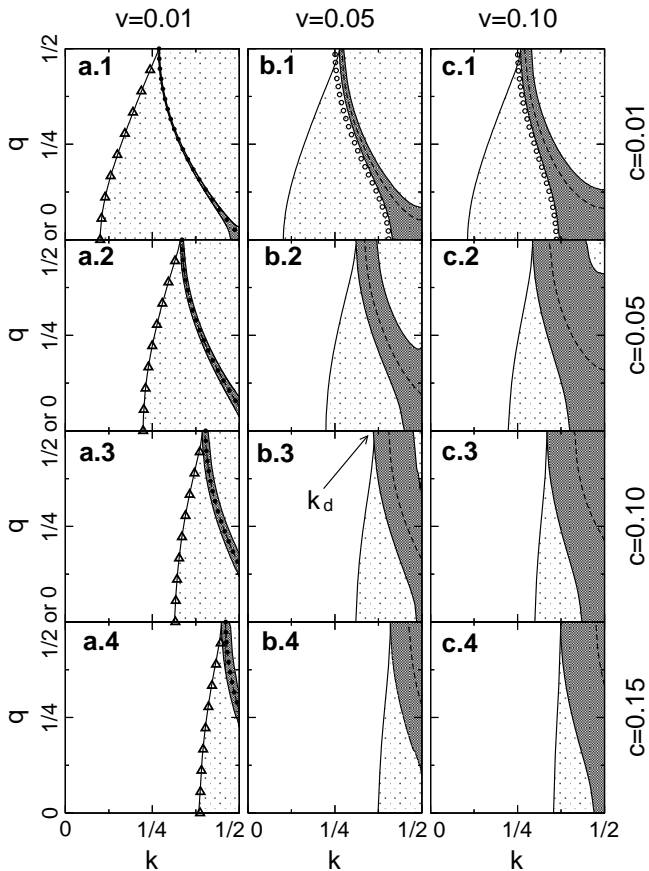}}
\end{center}
\caption{Stability phase diagram of BEC Bloch waves. $k$ is the wave number 
of BEC Bloch waves; $q$ denotes the wave number of perturbation modes.
In the shaded (light or dark) area, the perturbation mode has negative 
excitation energy; in the dark shaded area, the mode grows or decays 
exponentially in time.  The triangles in (a.1-a.4) represent 
the boundary, $q^2/4+c=k^2$, of saddle point regions at $v=0$.  The solid
dots in the first column are from the analytical results of Eq.(\ref{eq:kqc}).
The circles in (b.1) and (c.1) are based on the analytical expression
(\ref{eq:bb}). The dashed lines indicate the most unstable modes for 
each Bloch wave $\phi_k$.}
\label{fig:kq}
\end{figure}

The situation becomes more complicated in the case of $v>0$,
when the optical lattice is turned on. We resort to numerical
calculations to examine whether the matrices $M_k(q)$ are positive 
definite and the corresponding BEC Bloch waves are 
local energy minima. Our focus is on the Bloch states in the lowest
Bloch band, excluding the loop. 
The results are summarized in the stability phase diagrams in Fig.\ref{fig:kq}, 
where a wide range of values of $v$ and $c$ are considered in order to 
give a global picture of how the instabilities change with the two system 
parameters, $c$ and $v$. The results have reflection symmetry in $k$ and $q$, 
so we only show the parameter region, $0 \le k \le 1/2$ and 
$0 \le q \le 1/2$. 

In the shaded area (light or dark) of each panel of Fig.\ref{fig:kq}
the matrix $M_k(q)$ has negative eigenvalues, and the corresponding Bloch 
states $\phi_k$ are saddle points. For those values of $k$ outside the shaded 
area, the Bloch states are local energy minima and represent superflows.   
The superflow region expands with increasing atomic interaction $c$, and 
occupies the entire Brillouin zone for sufficiently large $c$.  
On the other hand, the lattice potential strength $v$ does not affect
the superflow region very much as we see in each row.
The phase boundaries for $v\ll 1$ are well reproduced from the analytical 
expression $k=\sqrt{q^2/4+c}$ for $v=0$, which is plotted as triangles
in the first column. 

The Landau instability of the states in the lower branch of the loop
has been discussed in Ref.\cite{smith2}. It is found that
the area of superfluidity increases when the interaction 
$c$ increases or the lattice strength $v$ decreases.

\subsection{Remarks on superfluidity}
So far, we have been identifying superfluidity as stability in the sense
of  Landau\cite{Lifshitz,fm}. In this approach,
there is an implicit assumption that the boson system has already
achieved phase coherence and can be described by a complex
order parameter, the macroscopic wave function. Whether
the state described by the macroscopic wave function 
possesses superfluidity is then examined by how the system 
responses to perturbations. With the assumption of phase coherence, 
this approach is not well-equipped to study quantum phase transitions 
at zero temperature, such as the superfluid-Mott-insulator 
transition\cite{mott_ex,mott_th}. Nevertheless, this approach can still 
provide some signals for the onset of some possible phase transition 
as we will discuss in the context of dynamical instability in 
the next section.

There are various other approaches to superfluidity.
One of particular interest was proposed by Fisher {\it et al.}
in Ref.\cite{fisher}, where superfluidity is defined as how strong 
the system is affected by an artificially imposed phase-twist 
at boundaries. Since only the superfluid part of the system responses 
to the phase-twist, by calculating the change of the system energy, 
one can find out the superfluid fraction of the Bose system. 
This approach has its root in the Josephson effect, where the 
phase coherence aspect of superfluidity was exposed for the 
first time\cite{pwa}. 

This phase-twist approach does not assume the phase coherence; to the 
opposite, it is particularly designed to calculate how much phase 
coherence a system has. It has been widely used in the Bose-Hubbard model to 
study the superfluid-Mott-insulator transition\cite{krauth}.
With the recognition that a BEC in an optical lattice can be
described by the Bose-Hubbard model\cite{mott_th}, this approach has 
now been adopted to study the superfluid-Mott-insulator transition 
for various systems of BECs in optical lattices\cite{rigol,roth}.

Both approaches are of perturbative nature, that is, they study
how the Bose system responses to small perturbations. In the Landau's
approach, all possible disturbances are considered; one is able
to determine the robustness of a superfluid as measured by the
critical velocity. In contrast, a special type of
perturbation is considered in the method of phase-twist, namely,
an artificial phase twist imposed at the boundaries. 
This perturbation is sensitive to the interplay between
kinetic energy and interaction. It is not clear
how this phase twisting corresponds to usual perturbations
(such as disorder or roughness) which would degrade the flow of
a normal fluid.  In short, the relation 
between these two pictures is still a subject of active debate\cite{burnett}.

\subsection{Dynamical instability}
One unique feature in the system of BECs in optical lattices is
dynamical instability, which does not exist in the absence
of either atomic interaction, $c=0$, or an optical lattice, $v=0$.
Many of the BEC Bloch waves can be dynamically unstable against 
certain perturbation modes $q$ only when both factors are present. By dynamical
instability, we mean that small deviations from a state
grow exponentially in the course of time evolution 
(see Fig.\ref{fig:landyn}).

The dynamical instability can be determined from the linear 
stability analysis of the GP equation Eq.(\ref{eq:nls}). Assume that the 
system experience a small disturbance $\delta\phi_k$ at a Bloch state,
\be
\psi(x,t)=e^{ikx-i\mu t}\Big[\phi_k(x)+\delta\phi_k(x,t)\Big]\,,
\ee
where the disturbance can be similarly written as
\be
\delta\phi_k(x,q) = u_k(x,q,t)e^{iqx}+v_{k}^*(x,q,t)e^{-iqx}\,.
\ee
Plugging the above wave function into Eq.(\ref{eq:nls}) and keeping
only the linear terms, we arrive at a linear dynamical equation,
\be\label{eq:first}
i{\partial\over \partial t}\pmatrix{u_{k} \cr v_{k}}=\sigma_z M_k(q)
\pmatrix{u_{k} \cr v_{k}},
\ee
where
\be
\sigma_z=\pmatrix{I&0\cr0&-I}\,.
\ee
The eigenvalues $\varepsilon_k(q)$ of the matrix $\sigma_z M_k(q)$ determine
the dynamical instability: If they are real for all $-1/2 \le q \le  1/2$,
the Bloch state is dynamically stable; if otherwise, that is, some of them
are complex, it is dynamically unstable. 

One who is familiar with the quasi-particle excitation in quantum fluid
may immediately notice that $\sigma_z M_k(q)$ also appears in the  
Bogoliubov approach and gives us the spectrum of the phonon 
excitation\cite{Lifshitz}. When all of its eigenvalues 
$\varepsilon_k(q)$ are real, the positive half are the phonon spectrum
while the other half are deemed non-physical.  However,
we will refer to the modes corresponding to the negative half as
anti-phonons for the sake of easy reference in the following discussion. 

Again, we first look at the case $v=0$, where the eigenvalues of 
$\sigma_z M_k(q)$ are
\be\label{eq:epsilon}
\varepsilon_{\pm}(q)=kq\pm\sqrt{q^2c+q^4/4}.
\ee
These eigenvalues are always real, implying that the BEC flows in a free 
space are always dynamically stable. When $v\neq 0$, the situation is totally 
different: the eigenvalues $\varepsilon_k(q)$ of $\sigma_z M_k(q)$ can be complex 
and Bloch states can be dynamically unstable. The results are summarized 
in Fig.\,\ref{fig:kq}, where these $\varepsilon_k(q)$ are complex in 
the dark-shaded areas. The dark shaded areas lie completely inside
the light shaded areas, as the result of a rigorous conclusion that only 
saddle-point Bloch states can have dynamical instability. 
Its proof is given in the appendix. 

The dynamical instability is the result of the resonance coupling
between a phonon mode and an anti-phonon mode by first-order
Bragg scattering. As shown in the appendix, the matrix $\sigma_z M_k(q)$
is real in the momentum representation, meaning its complex eigenvalue
can appear only in conjugate pairs and they must come from a pair of real
eigenvalues that are degenerate prior to the coupling. Degeneracies or resonances 
within the phonon spectrum or within the anti-phonon spectrum do not give rise to 
dynamical instability; they only generate gaps in the spectra.
Based on this general conclusion, we have analyzed dynamical instability 
for two extreme limits, $v\ll 1$ and $c\ll v$.

The limit, $v\ll 1$, corresponds to the cases shown in the first column of 
Fig.\,\ref{fig:kq}. For this limit, we can approximate the phonon spectrum 
and the anti-phonon spectrum with the ones given in Eq.(\ref{eq:epsilon}).
By equating them, $\varepsilon_{+}(q-1)=\varepsilon_{-}(q)$, for the degeneracy,
we find that the dynamical instability should occur on the following curves 
\be \label{eq:kqc}
k=\sqrt{q^2 c+q^4/4}+\sqrt{(q-1)^2 c+(q-1)^4/4}\,.
\ee
These curves are plotted as solid dots in Fig.\,\ref{fig:kq},
and they fall right in the middle of the thin dark-shaded areas. To
some extent, one can regard these thin dark-shaded areas as broadening
of the curves (\ref{eq:kqc}). It is noted in Ref.\cite{smith2} that
the relation (\ref{eq:kqc}) is also the result of
$\varepsilon_+(q-1)+\varepsilon_+(-q)=0$,
which involves only the physical phonons. Therefore, the physical meaning
of Eq.(\ref{eq:kqc}) is that one can excite a pair of phonons with total energy 
zero and with total momentum equal to a reciprocal wave number of the lattice.

The other limit, $c\ll v$, is the cases plotted the first row of 
Fig.\,\ref{fig:kq}. The open circles along
the left edges of these dark-shaded areas are given by
\be\label{eq:bb}
E_1(k+q)-E_1(k)=E_1(k)-E_1(k-q)
\ee
where  $E_1(k)$ is the lowest Bloch band of 
\be
H_0=-{1\over 2}{\partial^2\over \partial x^2}+v\cos (x).
\ee
In this linear periodic system, the excitation spectrum (phonon or anti-phonon) 
just corresponds to transitions from the Bloch states of energy $E_1(k)$
to other Bloch states of energy $E_n(k+q)$,  or vise versa.  
The above equation is just the resonance condition between such excitations in
the lowest band ($n=1$).  Alternatively, we can write the resonance 
condition as
\be
E_1(k)+E_1(k)=E_1(k+q)+E_1(k-q)\,.
\ee
So, this condition may be viewed as the energy and momentum conservation 
for two particles interacting and decaying into two 
different Bloch states $E_1(k+q)$ and $E_1(k-q)$.  This is
 the same physical picture behind Eq.(\ref{eq:kqc}).

One common feature of all the diagrams in Fig.\,\ref{fig:kq} is that
there is a critical Bloch wave number $k_d$ beyond which 
the Bloch waves $\phi_k$ are dynamically unstable.
The onset instability at $k_d$ always corresponds to $q=1/2$.
In other words, if we drive the Bloch state $\phi_k$ from $k=0$ to $k=1/2$
the first unstable mode appearing is always $q=\pm 1/2$, which 
represents period doubling.  Only for $k>k_d$ can longer wavelength
instabilities occur. The growth of these unstable modes drives the system 
far away from the Bloch state and spontaneously breaks the translational
symmetry of the system.  

The dynamical instability of the lower loop states is studied in 
Ref.\cite{smith2}. The range of $k$, where dynamical instability
occurs, is found decreasing with increasing $c$ and decreasing $v$.

\subsection{Tight-binding approximation}
\begin{figure}[!htb]
\begin{center}
\includegraphics[width=6.5cm]{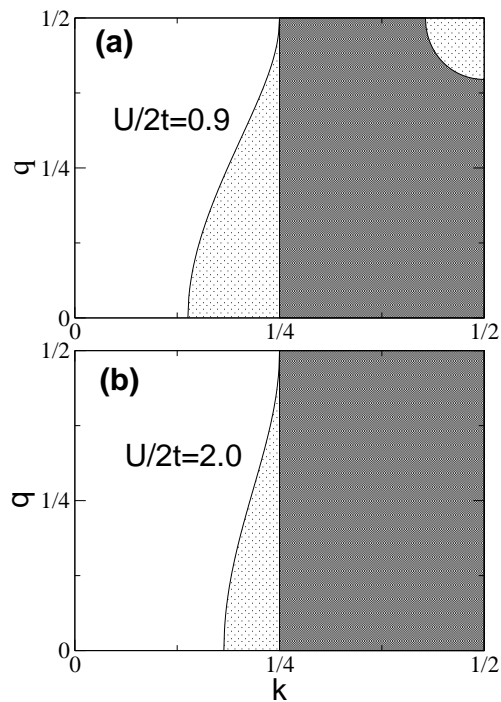}
\end{center}
\caption{Stability phase diagram in the tight-binding limit. Two different 
typical cases, $U/2t<1$ and $U/2t>1$, are shown in (a) and (b), respectively.
}
\label{fig:tight}
\end{figure}

We have so far presented a theory of how to determine the Landau 
and dynamical instabilities for the system of BECs in optical lattices.
It is valid for the full ranges of parameters $c$ and $v$ as long as the
mean-field treatment is adequate. It is, nevertheless, worthwhile to
consider the system in certain extreme limits, where
analytical results are available. The particular limit
that we now look into is the tight-binding limit, where the lattice
is so strong that the BEC system can be considered as a chain of trapped
BECs that are weakly linked. This limit was studied in Ref.\cite{smerzi}
with its focus on dynamical instability. In this case, the BEC trapped 
in each well of the lattice can be approximately described by
\be
\varphi_n(x)=\varphi^0(x-2n\pi)\,,
\ee 
where the real wave function $\varphi^0(x)$ is very localized 
in the well $[0,2\pi]$ such that
\be
\frac{1}{2\pi}\int dx\,\varphi_n(x)\varphi_m(x)\approx \delta_{mn}\,.
\ee
The tight-binding approximation is to write 
\be
\psi(x,t)=\sum_n W_n(t)\varphi_n(x)\,.
\ee
Substituting it into the GP equation (\ref{eq:nls}), we find after neglecting
a trivial energy constant\cite{smerzi},
\be
\label{eq:tight}
i\frac{\partial W_n}{\partial t}=-t(W_{n+1}+W_{n-1})+U|W_n|^2W_n\,,
\ee
where
\be
t=\frac{1}{2\pi}\int dx\,\Big(\frac{1}{2}\frac{\partial \varphi_n}{\partial x}
\frac{\partial \varphi_{n+1}}{\partial x}+\varphi_n\varphi_{n+1}v\cos(x)\Big)\,,
\ee
and
\be
U=\frac{c}{2\pi}\int dx\,\varphi_n^4(x)\,.
\ee
No surprise, the tight-binding equation (\ref{eq:tight}) has also 
a Bloch wave solution
\be
\label{eq:tbloch}
W^k_n=\exp{(i2kn\pi-\mu t)}
\ee
with $\mu(k)=U-2t\cos(2k\pi)$.
The instabilities of this Bloch wave solution can be determined by
following precisely the method described in the previous subsections.
We write down the perturbed state
\be
W_n(t)=W_n^k(t)+(ue^{2iqn\pi}+v^*e^{-2iqn\pi})e^{i2kn\pi-\mu t}\,.
\ee
which leads to an M matrix (similar to Eq.(\ref{eq:mmatrix})
\be
M_k(q)=\pmatrix{L(k+q)& U\cr U & L(-k+q)}.
\ee
In the matrix,
\be
L(k+q)=U+4t\sin(q\pi)\sin[(q+2k)\pi]\,.
\ee
By straightforward calculations, we find the eigenvalues of $M_k(q)$
\ba
\lambda_k(q)=U+4t\sin^2(q\pi)\cos(2k\pi)\pm\nonumber\\
\pm\sqrt{U^2+[2t\sin(2q\pi)\sin(2k\pi)]^2}\,,
\ea
and the Bogoliubov spectrum (eigenvalues of $\sigma_z M_k(q)$)
\ba
\varepsilon_k(q)=2t\sin(2q\pi)\sin(2k\pi)\pm
2\sin(q\pi)\times\nonumber\\
\times\sqrt{2tU\cos(2k\pi)+4t^2\sin^2(q\pi)\cos^2(2k\pi)
}\,.
\ea
Based on these two results, one can easily determine the stability phase
diagram of the Bloch wave solution (\ref{eq:tbloch}), which is 
plotted in Fig.\ref{fig:tight}. When $U/2t<1$, there is
a right border line for dynamical instability (see the top right corner 
of Fig.\ref{fig:tight}(a)); it is given by
\be
\sin^2(q\pi)=\frac{U}{2t|\cos(2k\pi)|}\,,~~~k>\frac{1}{4}.
\ee
This border disappears for $U/2t\ge 1$. For any value of $U/2t$, the 
left border of dynamical instability is at $k=1/4$ for any $q$ while
the boundary of Landau instability is described by
\be
\cos^2(q\pi)=\cos(2k\pi)\Big(\frac{U}{2t}+\cos(2k\pi)\Big)\,,~~k\le\frac{1}{4}.
\ee 
Note that the border line $k=1/4$ is consistent with our analysis
in the last subsection. It is the result of Eq.(\ref{eq:bb}) when
one uses the tight-binding band energy $E(k)\propto \cos(2k\pi)$.

It is evident that the phase diagram Fig.\ref{fig:tight} is a natural
extrapolation of the results shown in Fig.\ref{fig:kq}, and it also
follows the trend in Fig.\ref{fig:kq} as one increases the interaction:
the overall area of instability decreases while the dynamical instability
spreads more in the area of Landau instability.

\subsection{Experiments}
\begin{figure}[!htb]
\begin{center}
\resizebox*{8cm}{3.5cm}{\includegraphics*{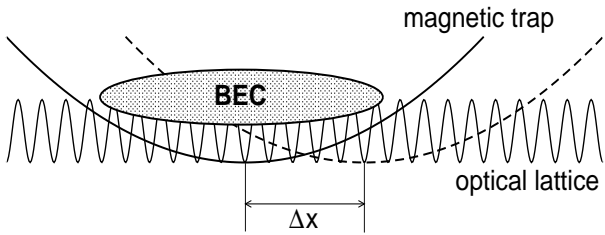}}
\end{center}
\caption{Schematic drawing of the experimental setup in Refs.\cite{burger,bishop}.}
\label{fig:exp}
\end{figure}

This dynamical instability  was observed in a recent experiment\cite{burger}, 
whose setup is schematically drawn in Fig.\ref{fig:exp}. A cigar-shaped BEC
was formed in a  magnetic trap superimposed with an 
optical lattice. This essentially prepared a BEC in a Bloch state at $k=0$,
the center of the Brillouin zone. The magnetic trap was kept on to keep 
the quasi-one dimensional configuration, but its center was suddenly 
shifted by $\Delta x$ along the longitudinal direction. This is equivalent 
to displacing the whole BEC off the center of the harmonic trap then
releasing it. 

According to the semi-classical theory\cite{Mermin}, 
\be\label{eq:semi}
\dot{x}={{\rm d}E(k)\over {\rm d}k}\,,\hspace{2cm}
\dot{k}=-\omega^2 x\,,
\ee
where $E(k)$ is the system energy and $\omega$ is the longitudinal frequency
of the magnetic trap. The BEC begins to oscillate in both the real
space and the momentum space (Brillouin zone). Since $E(k)\propto k^2$ 
near the center of the Brillouin zone, both oscillations 
are harmonic when  the displacement $\Delta x$ is small, as observed
in the experiment. However, when $\Delta x$ was over a certain 
critical value, say, $\Delta x_c$, it was observed that the oscillations
were disrupted and became dissipative.

This dissipative behavior is caused mainly by the dynamical instability in our opinion
although the authors of Ref.\cite{burger} attributed it to Landau instability.
When the displacement $\Delta x$ is small, the oscillation 
is confined near the center of the Brillouin zone, where
Bloch states are stable as shown in Fig.\,\ref{fig:kq}.
With the increase of $\Delta x$, the Bloch state will
be driven into the dangerous zone of dynamical instability; the
oscillation then can be disrupted by the growth of modes of instability.
In our units, this experiment has $v\sim 0.2$ and $c\sim 0.02$,
which is in the regime where the dynamical instability is rampant. 
Further experiments in Ref.\cite{burger} found that
$\Delta x_c$ increases with the decrease of the lattice strength $v$,
and there was no dissipation when the density of BEC was low.
These two observations basically rule out Landau instability as 
the main cause of dissipation. As we see in each row of 
Fig.\,\ref{fig:kq}, the Landau instability, as represented by the light-shaded
areas,  is affected very little by the change of $v$, thus $\Delta x_c$ should see no 
changes within experimental error when $v$ changes.
Also clear in Fig.\,\ref{fig:kq}, when $c$ is small, almost all Bloch 
states have Landau instability. As a result, more severe 
dissipation would have been observed for smaller $c$ if the Landau instability 
was the main contributer.

Very recently, there was another experiment with a similar setup, but
the optical lattice used is much stronger\cite{bishop}. The advantage of strong 
optical lattices is that dynamical instability is more severe, rendering Landau
instability less relevant. As a result, the interpretation of the experimental
observation is more definite and much less controversial than the one 
in Ref.\cite{burger}.

There are other possible ways to observe the dynamical instability.
One way is to prepare a Bloch state then monitor how 
its periodicity changes over with Bragg scattering of a probing laser beam
by the BEC cloud \cite{Phillips}. Another way is indirectly through
the breakdown of Bloch oscillations\cite{josephson,morsch,bo}. 
Within proper regime of parameters, the dynamical instability can be
very severe and thus drive the system far away from a Bloch state within 
a Bloch period. This leads to the breakdown of Bloch oscillations; the
observation of this breakdown is certainly an observation of dynamical instability.
However, one must be careful since the nonlinear Landau-Zener tunneling due to
the loop structure, as discussed in the last section,
can also leads to the destruction of Bloch oscillations. 
This ambiguity can be avoided by using dilute BEC such that 
$c<v$, where the loop does not exist.

\subsection{Discussion on dynamical instability}
As we emphasized above, dynamical instability happens only when both
the optical lattice and the interaction between atoms exist. However,
it does not imply that it is special to BECs in optical lattices.
To the contrary, it is very general. For example, it has been identified
in a BEC confined in a torus. In the following, we will demonstrate
this point with a one-mode problem. To provide a slightly different
angle, we will present it in the language of operator algebra and
use the Bogoliubov approach\cite{Lifshitz}. A discussion
on a similar two-mode problem can be found in Ref.\cite{anglin}.  

The one-mode problem is described by the Hamiltonian written as
\be
\label{eq:hgen}
H=A\hac\haa+{B\over 2}\hac\hac+{B\over 2}\haa\haa\,,
\ee 
where $A$ and $B$ are two positive real numbers. The two operator
$\hac$ and  $\haa$ are generation and annihilation operators, respectively, satisfying
the boson commutation relations
\be
[\hat{a}, \hat{a}^{\dagger}]=1\,,~~~~
 [\hat{a}, \hat{a}]=[\hat{a}^{\dagger}, \hat{a}^{\dagger}]=0\,.
\ee
This kind of quadratic Hamiltonians are often found as the result
of application of perturbation theory to a degenerate Bose gas near
a condensed state, such as Eq.(\ref{eq:approx}). For this simple one-mode case,
the $M$ matrix similar to Eq.(\ref{eq:mmatrix}) is
\be
M=\pmatrix{A&B\cr B&A}\,.
\ee

The Bogoliubov approach is to diagonalize the Hamiltonian (\ref{eq:hgen}) 
with the introduction of a new set of operators,
\be
\haa=u\hba+v^*\hbc~,~~~~~~~
\hac=u^*\hbc+v\hba
\ee
For $\hba$ and $\hbc$ to be boson operators, it requires that
\be\label{eq:boson}
|u|^2-|v|^2=1\,.
\ee
If $u$ and $v$ satisfy 
\be\label{eq:sm}
\sigma_z M\pmatrix{u\cr v}=\varepsilon\pmatrix{u\cr v}\,,
\ee
the Hamiltonian (\ref{eq:hgen}) becomes diagonal in $\hba$ and $\hbc$, up to
a constant,
\be\label{eq:dh}
H={\varepsilon+\varepsilon^*\over 2}\hbc\hba\,.
\ee

The phonon energy $\varepsilon$ is easily found by solving Eq.(\ref{eq:sm}),
and it is 
\be
\varepsilon=\pm\sqrt{A^2-B^2}\,.
\ee
When $A< B$, this energy becomes imaginary; we, again, encounter dynamical 
instability. The simplicity of this one-mode problem allows us look more 
carefully into this instability. We notice from Eq.(\ref{eq:dh}) that
$H=0$ when $\varepsilon$ becomes imaginary. At the same time,
we find with Eq.(\ref{eq:sm}) that $|u|^2-|v|^2=0$, violating the condition
(\ref{eq:boson}) for $\hba$ and $\hbc$ to be boson operators. 
In other
words, the Hamiltonian (\ref{eq:hgen}) defies the diagonalization by
the Bogoliubov approach when $A<B$.

The physics lies behind the failure of the Bogoliubov diagonalization
is the dynamical instability: the system is intrinsically unstable
and, therefore, does not have any phonon excitation. We can look into
this in a different representation. We write $\haa$ and $\hac$ in terms of
space and momentum operators,
\be
\haa={1\over\sqrt{2}}(\hat{x}+i\hat{p})~,~~~\hac={1\over\sqrt{2}}(\hat{x}-i\hat{p})\,.
\ee  
As a result, the Hamiltonian assumes the standard form of a harmonic
oscillator,
\be
H={A+B\over 2}\hat{x}^2+{A-B\over 2}\hat{p}^2\,.
\ee
It is evident here that, when $A<B$, the system is a harmonic oscillator with
a negative mass, an unstable system.

A quantum system with dynamical instability, such as a harmonic oscillator
with a negative mass, does not exist in nature. Its appearance in the
theoretical study only signals the onset of instability of the 
underlying unperturbed states. In our case, these unperturbed
states are the BEC Bloch states.  Once these Bloch states
develop dynamical instability, it means that the GP equation, which
predicts the existence of these Bloch states, is no longer a good
description of the system. The system enters a new unknown phase 
and demands a new mathematical description other than
the mean-field theory, the GP equation. In Ref.\cite{bishop}, the
new phase was called ``insulator''. There may be some experimental 
indications to call it ``insulator''; theoretically, we are yet to
develop a theory accurately depicting the new phase. Our theory, along
with the one in Ref.\cite{smerzi,bishop}, can
only predict when or where the dynamical instability sets in and
the system enters a new phase; it does not give any indication
how the new phase should look like and behave.

The dynamical instability is closely related to Landau instability. As already
briefly stated above, Bloch waves with dynamical instability must also have
Landau instability; on the other hand, Bloch waves with Landau instability
are not necessarily dynamically unstable. This comes from a rigorous
result on the relation between eigenvalues of $M_k(q)$ and $\sigma_z M_k(q)$
as proved in the Appendix: when all the eigenvalues of $M_k(q)$ are all positive,
the eigenvalues $\varepsilon_k(q)$ of $\sigma_z M_k(q)$ are all real.

\section{Summary}
In summary, we have examined the superfluidity and many of its related 
properties of BECs in optical lattices. We studied the tunneling between 
Bloch bands and found that the tunneling is not zero even in the adiabatic 
limit when the repulsive interaction $c$ is strong enough.
This dramatic tunneling behavior is due to the loop structure in the Bloch band,
a result of strong superfluidity. 
We investigated Landau instability and dynamical instability 
of the system. With a phase diagram, we showed where Landau and dynamical instability
occur and how they change with the variation of the two system parameters,
$c$ and $v$. We have discussed the experimental observation of dynamical
instability and its implication in this system.

\acknowledgments{
This work is supported by NSF and the Welch Foundation.
We thank Jie Liu and R.B. Diener for their collaborations
in the work reviewed here. BW is grateful for insightful discussions
on dynamical instability with Junren Shi and J.R. Anglin}

\appendix
\section{Mathematical Properties of $M_k(q)$ and $\sigma_z M_k(q)$}
For simplicity, we will drop subscript $k$ and refer to these two matrices simply
as $M(q)$ and $\sigma_z M(q)$ in this appendix. We are interested in the properties
of their eigenvalues as defined in 
$$
M(q)X(q)=\lambda(q)X(q),\eqno(A.1)
$$
and 
$$
\sigma_z M(q)Y(q)=\varepsilon(q) Y(q)\,.\eqno(A.2)
$$
The eigenvalues $\lambda(q)$ are all real since the matrix $M(q)$ is Hermitian. 
On the other hand, the matrix $\sigma_z M(q)$ is not Hermitian and 
its eigenvalues $\varepsilon(q)$ are not necessarily all real. 
However, due to its special structure, the eigenvalues $\varepsilon(q)$ 
enjoy some very interesting properties. There is also a very important
relation between the two sets eigenvalues, $\lambda(q)$ and $\varepsilon(q)$.\\

\noindent (i){\it 
If $\varepsilon$ is one eigenvalue, then $\varepsilon^*$ is
also an eigenvalue of $\sigma_z M(q)$. In other words, if the matrix $\sigma_z M(q)$ 
has complex eigenvalues, they appear in conjugate pairs.} \\

This becomes evident when we write the matrix in a form, where all its elements
are real. Note that, similar to the Hamiltonian operator
in the left hand side of Eq.(\ref{eq:nlsi}), $\sigma_z M(q)$ is periodic in space so that one
can express it in the space of Fourier series. The expression is
$$
\sigma_z M(q)=\pmatrix{A & B\cr -B^{T} & -C}\,, \eqno(A.3)
$$
where 
$$
A_{mn}={1\over 2\pi}\int_0^{2\pi}dx e^{-imx}{\mathcal L}(k+q)e^{inx}\,,\eqno(A.4)
$$
$$
B_{mn}={c\over 2\pi}\int_0^{2\pi}dx e^{-imx}\phi_k^2(x)e^{inx}~~~~~~~\,,\eqno(A.5)
$$
and
$$
C_{mn}={1\over 2\pi}\int_0^{2\pi}dx e^{-imx}{\mathcal L}(-k+q)e^{inx}\,.\eqno(A.6)
$$
One can check by straightforward calculations that all the elements, $A_{mn}$,
$B_{mn}$, and $C_{mn}$ are real. This particular expression is very
convenient for numerical calculation: after neglecting Fourier terms of orders
higher than a cut-off order, say, $N$, one reduces $\sigma_z M(q)$ to a real 
$(4N+2)\times(4N+2)$ matrix. To our experience, $N=10$ is sufficient.\\

\noindent (ii) {\it If $\varepsilon$ is an eigenvalue of $\sigma_z M(q)$, then 
$-\varepsilon^*$ is an eigenvalue of $\sigma_z M(-q)$. }\\

Write Eq.(A.2) in another form,
$$
\pmatrix{{\mathcal L}(k+q) & c\phi_k^2\cr
-c\phi_k^{*2}& -{\mathcal L}(-k+q)}\pmatrix{u\cr v}=\varepsilon\pmatrix{u\cr v}\,.
\eqno(A.7)
$$

By some simple algebraic manipulations, one immediately observes that
$-\varepsilon^*$ is an eigenvalue of $\sigma_z M(-q)$ with the corresponding 
eigenvector being $\{v^*, u^*\}$.  \\

\noindent (iii) {\it 
When all the eigenvalues $\lambda(q)$ are positive,
the eigenvalues $\varepsilon(q)$ must be all real.}. \\

We multiply both sides of Eq.(A.2) by $\sigma_z$ then by $Y^{\dagger}$,
and obtain
$$
Y^{\dagger}M(q)Y=\varepsilon(q)Y^{\dagger}\sigma_z Y\,.\eqno(A.8)
$$
Since the matrix $M(q)$ is Hermitian, the left hand side of the above equation
is always real. As a result, if an eigenvalue  $\varepsilon(q)$ is complex, then
for its corresponding eigenvector $Y$,
one has $Y^{\dagger}M(q)Y=Y^{\dagger}\sigma_z Y=0$. 

When all eigenvalues of $M(q)$ are positive, the left hand side of Eq.(A.8) 
must be positive. In this particular case, one can conclude that
the eigenvalues $\varepsilon(q)$ are all real and share the sign of 
$Y^{\dagger}\sigma_z Y$.

\end{document}